\documentclass[12pt]{article}
\usepackage{mathptmx}

\usepackage[utf8]{inputenc}
\usepackage{graphicx}
\usepackage{amssymb}

\usepackage{natbib}
\usepackage{hyperref}

\usepackage[letterpaper]{geometry}
\usepackage{enumitem}
\setlist[itemize]{noitemsep}

\def\p{}
\def\pp{}

\begin{document}

\title{Empirical structure physicalism and realism, Hempel's dilemma, and an optimistic meta-induction}
\author{{Bal\'azs Gyenis}, {\em Synthese} (2025) {\bf 206}, 76. \\ \\ Institute of Philosophy, Research Centre for the Humanities, Budapest \\  email: {gyepi@hps.elte.hu} \\ ~}
\date{Pre-typesetting, accepted version. Published version:  \\
ePDF: \url{https://rdcu.be/exUuE}  \\
DOI: \url{https://doi.org/10.1007/s11229-025-05160-x} }

\maketitle
\thispagestyle{empty}

\begin{abstract}
Motivated by a generalization of Hempel's dilemma, I introduce a novel notion of empirical structure, as well as theory supervenience as a new reductive relationship between theories. One theory supervenes on another theory if the empirical structure of the latter theory refines the empirical structure of the former theory. I then argue that (1) empirical structure physicalism, the thesis that the current special sciences supervene both on current and on future physics, avoids both horns of Hempel's dilemma; (2) in particular, mental theories remain empirically dispensable in the future; (3) empirical structure realism, the thesis that earlier theories of physics supervene on later theories of physics, is supported by an optimistic meta-induction; (4) this optimistic meta-induction can coexist with the well-known pessimistic meta-induction; (5) empirical structure physicalism is appropriately labeled as a type of physicalism; and (6) empirical structure physicalism is compatible with multiple realization. To illustrate the plausibility of empirical structure physicalism, I also briefly address the so-called knowledge argument.
\end{abstract}

%\clearpage
%\pagenumbering{arabic} 

\section{Introduction}

Philosophers of mind call Hempel's dilemma an argument by \citep{Crane-Mellor1990,Melnyk1997} against {\em metaphysical physicalism}, the thesis that everything that exists is either `physical' or ultimately depends on the `physical'. Their argument is understood as a challenge to the idea of fixing what is `physical' by appealing to a theory of physics. The dilemma briefly goes as follows. On the one hand, if we choose a current theory of physics to fix what is `physical', then, since our current theories of physics are very likely incomplete, the so-articulated metaphysical physicalism is very likely false. On the other hand, if we choose a future theory of physics to fix what is `physical', then, since future theories of physics are currently unknown, the so-articulated metaphysical physicalism has indeterminate meaning. Thus, it seems we can rely neither on current nor on future theories of physics to satisfactorily articulate metaphysical physicalism. Recently \citep{Firt-Hemmo-Shenker2022} argued that the dilemma extends to any theory that gives a deep-structure and changeable account of experience \citep[including dualistic theories, although cf.][]{Buzaglo2024}.

Even though the dilemma was christened after Carl Hempel, his original argument was neither mounted against a metaphysical thesis nor was understood as a challenge to fixing what is `physical' by choosing a theory of physics. Hempel's original dilemma was directed against the ``physicalistic claim that the language of physics can serve as a unitary language of science'' \citet[\pp 194--195]{Hempel1980}.  Thus, Hempel's {\em linguistic physicalism} states a connection between the language of physics and the languages of the special sciences, and it invokes neither a direct claim about what exists, nor employs the term `physical' which would in turn require further clarification. In his argument Hempel asked which theory of physics is meant by the thesis of linguistic physicalism and pointed out that neither current theories of physics seem to be able to do the job (due to their changeable character), nor a theory that would be defined through a prior notion of `physical phenomena' (due to the inherent unclarity of the notion).\footnote{Hempel's primary worry was thus not how to fix the meaning of `physical' by invoking a theory of physics; on the contrary, he only briefly considered --- and dismissed --- invoking the notion of `physical phenomena' to fix a theory of physics, and he did so only in one of the horns of his dilemma. For more on his aversion against the notion of `physical phenomena', see also \citep[181]{Hempel1969}.}

The metaphysical and the linguistic dilemmas differ in the type of reductive relationship posited between physics and the special sciences (broadly construed as to include mental theories): theory-oriented metaphysical physicalism addresses the reductive relationship between the entities posited by physics and the entities posited by the special sciences, while linguistic physicalism addresses the reductive relationship between the language of physics and the language of the special sciences. The similarity of the two dilemmas lies in that that they express similar problems with identifying physics with either a current theory of physics that is likely to undergo future change, or with a future or ideal theory whose content is currently indeterminate or circularly defined.

Hempel's dilemma can thus be understood not merely as a particular threat against theory-oriented metaphysical physicalism, but as a general threat against any precise articulation of a reductive relationship between physics and the special sciences. In this article I understand Hempel's dilemma in this general way. Instead of adhering to physicalism as a metaphysical (or linguistic) thesis, one of my goals is to seek a type of reductive relationship between physics and the special sciences which can avoid the challenges of both horns of Hempel's dilemma, while remaining informative enough to merit being called a type of physicalism. 

In this article I propose such a novel reductive relationship which I call empirical structure physicalism. To do so, after a brief discussion of Hempel's dilemma in Sect. \ref{section_hempelsdilemma} I introduce a precise notion of empirical content (\ref{subsection_empiricalcontent}) and empirical structure (\ref{subsection_empiricalstructure}). Succinctly and with simplification, the {\em empirical content of a space-time region according to a theory} is the set of those empirically adequate models of the theory which can be understood as representing certain observable aspects of happenings throughout and restricted only to the space-time region; the {\em empirical structure of a theory} is the partition composed of the sets of all space-time regions which have the same empirical content according to the theory. One theory then supervenes on another theory if the empirical structure of the latter theory refines the empirical structure of the former theory (\ref{subsection_theorysupervenience}). Given these notions I formulate various {\em a posteriori} supervenience theses between theories which are connected in a diachronic (past, present, and future theories of physics) or in a synchronic (the special sciences and physics) way (\ref{subsection_theorysuperveniencephysicalism}). 

Empirical structure physicalism claims that the current special sciences supervene both on current and on future physics. Empirical structure physicalism is a simple conjunction of its currentist and futurist varieties. On the one hand, I argue that currentist empirical structure physicalism can hold even if current physics were incomplete, and thus can avoid the first horn of Hempel's dilemma. On the other hand, I point out that futurist empirical structure physicalism is a simple deductive consequence of currentist empirical structure physicalism and future empirical structure realism (the thesis that current theories of physics supervene on future theories of physics) due to transitivity of the theory supervenience relation. Thus, if one could further argue for future empirical structure realism then the threat of future changes in physics would also be avoided; in particular mental theories would remain empirically dispensable in the future (\ref{section_mainarguments}). I complete the argument by showing that an optimistic meta-induction strongly supports future empirical structure realism (\ref{section_optimisticmetainduction}). 

Sect. \ref{subsection_physicalisms} surveys requirements imposed on a desirable notion of physicalism in the contemporary philosophical literature and argues that empirical structure physicalism satisfies almost all of these desiderata. In Sect. \ref{subsection_dualisms} I discuss the relationship of empirical structure physicalism and dualism, while in Sect. \ref{subsection_multiplerealization} I address its compatibility with multiple realization. Sect. \ref{section_conclusion} adds closing remarks.

\section{Hempel's dilemma}\label{section_hempelsdilemma}

Hempel's dilemma against metaphysical physicalism can be seen as a contradiction among four premises (P1)-(P4):

(P1) {\em Metaphysical physicalism}: everything that exists is either `physical' or ultimately depends on the `physical'. 

\noindent Metaphysical physicalism is intended as a non-tautological thesis with a determinate meaning, and the distinction made by (P1) is meant to be non-trivial: for instance, mental entities are usually not considered `physical' entities themselves, rather ultimately dependent on `physical' entities.

To make sense of (P1), we need to define the term `physical'. Intuitively, physics tells us what `physical' is: a theory of physics suggests the kind of `physical' things that exist. However, different theories of physics may suggest the existence of different `physical' things. This intuition motivates theory-oriented articulations of (P1): 

(P2) To fix the meaning of the thesis of physicalism employed by (P1), we need to fix a theory or theories of physics.

So the question is: which theories of physics should we choose in premise (P2)? It seems we can either choose current, future, or ideal theories of physics. However, each option has its problems.

The main problem with choosing current theories of physics in (P2) is that

(P3) It is very likely that current theories of physics do not give a complete description of reality. 

\noindent In particular, the well-known pessimistic meta-induction argument suggests that it is very likely that future, more complete, and more likely true theories of physics will posit elements of reality whose existence is neither suggested by current theories of physics, nor ultimately dependent upon things whose existence is suggested by current theories of physics. But then, it is very likely false that everything that exists is either a thing whose existence is suggested by current theories of physics or is ultimately dependent upon things whose existence is suggested by current theories of physics. Thus, if we chose current theories of physics in (P2), (P1) would be very likely false.

The main problem with choosing future theories of physics in (P2) is that

(P4) Future theories of physics are currently unknown. 

\noindent Since we don't know what kind of things exist according to a currently unknown theory, if we chose a future theory of physics in (P2), the meaning of (P1) is currently indeterminate. Although the meaning of (P1) would become determinate once the chosen future theory of physics becomes known, we have no assurance that this future meaning will align with its current intent. To see this, note that (P1) is meant to be non-trivial: (P1) intends to maintain a distinction between entities that are `physical' and entities which are not `physical' but ultimately dependent on the `physical'. In particular, (P1) intends to say that mental entities are themselves not among the `physical' entities, only ultimately dependent upon `physical' entities. However, since future theories of physics are currently unknown, it seems possible that a future theory of physics will claim that mental entities are among the `physical' entities. Such a claim would then trivialize, and hence substantially alter, the current intent of (P1).

Finally, if by an ideal theory of physics we mean a theory of physics which, by definition of being ideal, accounts for all things, then by choosing this ideal theory in (P2), (P1) would become tautological (since everything that exists then, by definition, becomes a thing whose existence is suggested by the ideal theory).

In sum, it follows that (P1) is either very likely false, or its meaning is currently indeterminate with no assurance that its future determinate meaning will align with its current intent, or it is tautological. 

Premises (P1)-(P4) thus form a set of inconsistent but `epistemically attractive' propositions, that is, a {\em philosophical problem} in the sense of \citep[78]{Tozser2023}; a {\em solution} of such a philosophical problem is another, but consistent set of propositions which respect all defensible epistemic attractions behind the original inconsistent set of propositions \citep[170--171]{Gyenis2023}. 

Philosophers who insist that the only philosophically interesting way to define physicalism is (P1) and who find physicalism and (P3)-(P4) epistemically attractive attempt to replace premise (P2) and define the term `physical' in a non-theory-oriented way. In doing so, metaphysicians often rely on concepts such as `paradigmatic physical object', `paradigmatic physical effect', `family resemblance', `natural kind', `fundamentally mental entity', etc. \citep[for details, see][]{Horzer2020}. I do not engage with these solution attempts here, but merely remark that I find attempts to divest the notion of physicalism from clearly articulated theories of physics --- and to articulate physicalism instead in terms of less-than-ideally clear concepts --- epistemically much less attractive than replacing one type of reductive relationship between physics and the special sciences (which I already find suspect for independent reasons discussed later) with another.

To solve a philosophical problem, any proposition from the inconsistent set could be a target for change; in the case of Hempel's dilemma, this includes premise (P1). We have already considered another way to understand physicalism: Hempel's linguistic physicalism (P1'). Unfortunately, (P1') offers no help here: as we have seen in the Introduction, \citet[\pp 194--195]{Hempel1980} argues that premises (P1'), (P2)-(P4) lead to a contradiction, with a reasoning similar to the one reconstructed above. Thus, replacing metaphysical physicalism (P1) with linguistic physicalism (P1') does not solve the philosophical problem.

Hence, it remains an open question whether one could solve Hempel's dilemma by keeping premises (P2)-(P4), and hence by replacing premise (P1) with another reductive relationship between physics and the special sciences, one which respects all of our defensible epistemic attractions behind (P1)-(P4). In this essay I propose that empirical structure physicalism is a good candidate for this role.

\section{Empirical structure and theory supervenience}\label{section_theorysupervenience}

In this section, I develop the notions of a {\em history-description} and an {\em empirical adequacy relation}; the former can be understood as a generalization of a model of the semantic view of theories, and the latter as a generalization of van Fraassen's empirical adequacy. Other important concepts --- including that of {\em formal content}, {\em empirical content}, {\em empirical structure}, {\em empirical theory}, {\em theory supervenience}, {\em empirically dispensable}, and {\em non-fundamental} --- and theses are at a certain point explicated using these two basic notions and the concept of a space-time region.

\subsection{History-description, empirical adequacy, empirical content}\label{subsection_empiricalcontent}

Recall that van Fraassen defines a theory to be empirically adequate if it has at least one model whose empirical substructures are isomorphic to the structures which are described in experimental and measurement reports of the observable phenomena \citep[64]{vanFraassen1980b}. Thus, for van Fraassen a theory specifies a set of models, and in addition, the theory specifies what the empirical substructures of these models are; empirical adequacy of a theory is then tied to the question whether there is at least one model whose empirical substructures fit the experimental and measurement reports of the observable phenomena. 

Thus first, it is clear that van Fraassen's notion of empirical adequacy as a relation between a theory and the phenomena is derivative from a relation between a model and the phenomena. Second, it is also clear that van Fraassen does not identify a theory simply with a set of models (as does his starting point, the semantic view):\footnote{\label{footnote_repr}
To understand the nature of scientific theories, one must address the following questions: (1) what is a theory? (2) what are the representational vehicles a theory uses to describe the world? and (3) under what conditions can these vehicles fulfill their representational role? The first two questions are frequently answered by identifying a theory and a representational vehicle with certain types of formal constructions, and one focal point of the philosophical debate is to say what these formal constructions are. Thus, early variants of the so-called syntactic (or linguistic) view identify a theory with a set of sentences of a formal language and identify the representational vehicles with certain sentences which are logically compatible with the so-understood theory. The so-called semantic (or model-theoretic) view identifies a theory with a set of model theoretical structures and identifies the representational vehicles with subsets of the same structures. Dual views offer a mixed reply to the first two questions (a theory is a set of sentences, while representational vehicles are the logical models of these sentences), and there are other formally inclined views which are difficult to categorize along these lines. For instance, many philosophers of physics tend to identify a theory of physics with differential equations expressing laws and representational vehicles with solutions of these differential equations, but try to distance themselves from making any syntactic or model theoretical commitments. For a good overview of options, see \citet{Frigg2022}.

In the aforementioned accounts, once we identify a formal construction for a theory and another for its representational vehicles, there is a direct link between a particular theory and the set of representational vehicles of the theory. The direct link is established by the notion of logical (or nomic) possibility: the representational vehicles of a particular theory are simply formal constructions which are, in some precise sense, logically compatible with (the laws of) the theory. For instance, the representational vehicles are either certain sentences which don't contradict the sentences of the theory (early syntactic view), or models which satisfy the sentences of the theory (dual view), or subsets of the theory (semantic view), or trajectories which are solutions of the differential equations of the theory, etc. The direct link also works from the other direction: instead of starting with particular theory, we can start by fixing a set of representational vehicles and then ask what is the theory they are representational vehicles of; for the semantic view the answer is an immediate identification of the theory with the set of representational vehicles, while for the early syntactic view or for dual views this question raises interesting technical issues about axiomatizability vs. characterizability of a class of models which have preoccupied many logicians of the past century.

However, there are several reasons to doubt the direct identification of a scientific theory with its set of representational vehicles. \citep{Norton2022} and others argued that possibility in science is best understood as an inductive concept which is distinct from mere logical compatibility (and nomic possibility, in Norton's terminology, is mere logical compatibility, see \citealt{Gyenis2025}). Pragmatic views emphasize non-formal features of scientific theories \citep[see][]{Winther2020}, and such non-formal features cannot be parts of an identity-criterion for theories if we simply identify a theory with its set of formal representational vehicles. In general, although logical analysis is useful as a tool for elucidating the formal content of a scientific theory, a scientific theory is not merely a formal, but also an empirical theory: it also has empirical content. For instance, according to a scientific theory, some of the logically possible representational vehicles of the theory are empirically adequate to a particular phenomena at a particular place and a particular time, while others are not. It is unclear how one could simply identify a scientific theory with all of its logically possible representational vehicles without saying anything about the circumstances under which some representational vehicles of the theory are empirically adequate, that is, how the third question above relates to the first two.
}
in addition to specifying its models, a theory is something that also specifies what the empirical substructures of a model are. To wit:
\begin{quote}
To present a theory is to specify a family of structures, its models; and secondly, to specify certain parts of those models (the empirical substructures) as candidates for the direct representation of observable phenomena. \citep[64]{vanFraassen1980b}
\end{quote}

There are a few problems with van Fraassen's notion of empirical adequacy. If we required the empirical substructures of a model of a theory to (1) be able to represent all experimental and measurement reports of all observable phenomena (2) with an exact fit (3) throughout our entire universe, then no current theory of physics would be empirically adequate. This is so because we currently lack a universal theory of physics which would be capable to describe all currently observable phenomena. Hence (1) no current theory is capable to describe all observables, (2) no current theory provides exact fit for even a select type of observables, only fit with a certain degree of approximation based on neglecting effects which are not described by the theory, and (3) even those current theories which provide fit with a very high degree of approximation for a select type of observables, for any set degree of approximation they can only provide predictive fit within the set degree of approximation for certain places and for a certain amount of time, in short, only for certain {\em space-time regions}. Thus, it seems clear that if we wanted empirical adequacy to be a notion that is applicable to our current scientific theories, then we need to relativize fit to (1) certain aspects \citep[747]{Hemmo-Shenker2022} of the world, (2) certain levels of approximation, and (3) certain space-time regions. In short, we need to allow for idealization, approximation, and localization. The last condition requires models of a theory to be understood as representing happenings (events, properties, entities, states, beables, etc.) throughout and restricted to certain space-time regions --- in agreement with scientific practice where representational vehicles are almost never understood as representing our entire actual world.

I intend to remain neutral regarding which type of construction best characterizes the representational vehicles of scientific theories. I use the collective term {\em history-description} for any account of how happenings may unfold in some parts of the world for some time. Thus, in the terminology of \citep[48]{Frigg2022}, a history-description is a representational model whose target system comprises certain aspects of happenings in a space-time region of our actual world. As far as scientific theories are concerned (for a more general case, see Sect. \ref{subsection_empiricalstructure}), Readers may freely substitute their preferred account for representational vehicles (certain sets of sentences, models of certain sentences, solutions of certain differential equations etc., see footnote \ref{footnote_repr}) for my term history-description, or may also take a sociologically inclined view and define the set of history-descriptions of a theory by assuming that a competent user of a scientific theory would readily recognize a history-description of the theory as such. The next subsection illustrates the notion through several examples.

With these preliminary remarks in mind, I define that a {\em history-description $h$ of a theory $T$ is empirically adequate to a space-time region $R$ according to $T$} if, given certain selected observable aspects of happenings, (i) according to $T$ it is admissible for $h$ to represent the selected observable aspects and to not represent other observable aspects of happenings (``idealization''), (ii) $h$ can be understood to represent the selected observable aspects of happenings throughout $R$ and restricted only to $R$ (``localization''), and (iii) what the so-understood $h$ says about observable quantities approximately matches, with an approximation allowed by $T$, the experimental and measurement reports of the selected aspects of happenings in $R$ (``approximate truth about selected observable aspects''); in short, if the history-description saves some selected observable phenomena in the space-time region with a certain approximation. Thus, empirical adequacy to a space-time region depends on what aspects of happenings the history-description aims to represent (what idealizations are admissible), whether a history-description can be understood to represent aspects of happenings throughout and restricted to the space-time region, what the history-description says about observable quantities in said region, and what counts as an admissible level of approximation for fit with measurement reports, according to the theory. I add further qualifications in Sect. \ref{section_optimisticmetainduction} regarding use of common sense statistical techniques guarding against cognitive bias and in Sect. \ref{section_conclusion} regarding the {\em prima facie} theory-dependent term `space-time region', but beyond these remarks I treat the notion of empirical adequacy as primitive.

It should be clear that the so-defined empirical adequacy of a history-description $h$ to space-time region $R$ according to $T$ is a very different concept than wholesale truth of $h$ about $R$. My notion of empirical adequacy only requires approximate truth of $h$ regarding certain observable aspects of happenings in $R$. A history-description $h$ can be empirically adequate even if none of its non-observable claims (or none of its claims about observable aspects that are not of interest to the theory $T$) about $R$ are true.

Empirical adequacy, construed as a relation between a history-description and a space-time region, naturally leads to a notion of empirical content. I define the {\em empirical content $e^T_R$ of a space-time region $R$ according to a theory $T$} as the set of those history-descriptions of $T$ which are empirically adequate to $R$ according to $T$. Formulaically, if $T$ is a theory, $H^T$ is the set of all history-descriptions of $T$, $R$ is a space-time region of our actual world $M$, and $EA^T(h,R)$ is the above-defined empirical adequacy relation, then the empirical content of $R$ according to $T$ is $e^T_R  ~ \doteq ~ \{ h \in H^T ~|~  EA^T(h,R) \}$. The {\em empirical content of a theory} is then, extensionally, a map that assigns to every space-time region a set of history-descriptions; intensionally, this latter set consists of all those history-descriptions of the theory which are empirically adequate to the space-time region according to the theory.\footnote{
Although \citet{vanFraassen1980b} never explicitly defines what he means by empirical content, from context (see esp. pages 45 and 47) it is clear that his notion of the empirical content of a theory $T$ corresponds to my notion of the empirical content $e^T_M$ of our entire actual world $M$ in case when theory $T$ is such that it only allows for exact fit with all observable phenomena. My notion of empirical content of a theory $T$, in contrast, is a map $R \mapsto e^T_R$ for all $R \subseteq M$, and is also applicable for theories that explicitly allow for approximate fit with observable phenomena of some aspects of the world. My notion thus can be seen as a practical purpose and space-time region relative generalization of van Fraassen's. 
}

\subsection{History-description, empirical adequacy, empirical content: examples}\label{subsection_empiricalcontentexample}

Newtonian physics can describe the motion of the planets in our Solar System over a span of ten years (say, space-time region $R_{10}$). The $H^{N}$ set of all history-descriptions of Newtonian physics consists of solutions of initial value problems of Newton's laws of motion. Consider four trajectories that solve the following Newtonian initial value problems: (a) two point particles, one with a very large mass initially at rest the center of the coordinate system, another with a very small mass, initially displaced from the center with a non-zero velocity pointing in any direction other than toward the first particle; (b) ten point particles, or (c) ten solid spheres, or (d) $10^{55}$ perfectly elastic spheres arranged in a way so that they form ten spherical clouds, in cases (b)--(d) with respective masses, initial positions and velocities matching those of the Sun and the nine planets. Choose four trajectory segments from these trajectories with a length so that they can represent one object revolving around another exactly ten times. These four trajectory segments are then history-descriptions which can represent aspects of the happenings throughout and restricted only to $R_{10}$: the trajectory segment corresponding to initial value (a) can represent the motions of the centers of mass of the Sun and the Earth; to (b) can represent the same also for eight additional planets; to (c) can also represent, in addition, the extensions of the planets; to (d) can even represent the atomic constituents of the planets (although not in an empirically adequate way) for ten years. According to Newtonian physics, the idealization that a history-description only represents the motions of certain planets is admissible, and the level of approximation which was available for astronomers in the 19th century is admissible for Newtonian physics. Given such pragmatic choices, and after suitable coordination of geometry with physical space and appropriate choices of measurement units, these four trajectory segments all turn out to be empirically adequate to $R_{10}$, and hence they all belong to the empirical content $e^{N}_{R_{10}}$.

However, longer trajectory segments of the same initial value problems would fail to be empirically adequate to the motion of the planets in our Solar System over a span of a million years (say, space-time region $R_{10^6}$). The perihelion precession of Mercury shows 43 arcseconds of discrepancy per century with Newtonian predictions even at 19th century measurement precisions, and the cumulative discrepancy over a million years becomes so large that Newtonian descriptions of planetary motion do not meet any level of approximation admissible for Newtonian physics. Hence neither of the above four type of history-descriptions would be empirically adequate to $R_{10^6}$; and this is also the case for any other history-descriptions of Newtonian physics, no matter how detailed, since ignoring representing the positions and velocities of objects is not an admissible idealization according to the theory. Hence the empirical content $e^{N}_{R_{10^6}}$ of the space-time region $R_{10^6}$ is {\em empty} according to Newtonian physics. The same holds for certain very small space-time regions (say, space-time region $R_{10^{-20}}$): due to quantum physical effects their empirical content is also empty according to Newtonian physics. 

Thus, in my account the space-time regions which have non-empty empirical content according to a theory characterize the {\em domain of empirical adequacy of the theory}. In contrast, the space-time regions whose empirical content equals the set of all history-descriptions of the theory characterizes the circumstances in which the theory is {\em unfalsifiable}. When $e^T_R \neq \emptyset$ and $e^T_R \neq H^T$ the empirical content is non-trivial: the `narrower' $e^T_R$ is the more `risky' claim the theory makes about $R$; in tandem, the `wider' $H^T \setminus e^T_R$ is the more `potential falsifiers' the theory has for $R$.

The same space-time regions typically have different empirical contents according to different theories. As we have seen, the empirical content of the aforementioned three space-time regions $R_{10^{-20}}$, $R_{10}$ and $R_{10^{6}}$ according to Newtonian physics are $e^{N}_{R_{10^{-20}}} = \emptyset$, $e^{N}_{R_{10}} \neq \emptyset$, and $e^{N}_{R_{10^{6}}} = \emptyset$. General Relativity (as a theory whose history-descriptions satisfy Einstein's field equations) is empirically adequate on the longer time scale, while still fails to be empirically adequate on the short one, thus $e^{GR}_{R_{10^{-20}}} = \emptyset$, $e^{GR}_{R_{10}} \neq \emptyset$, and $e^{GR}_{R_{10^6}} \neq \emptyset$. For Quantum Mechanics, which does not yet have an extension to curved space-times, we have $e^{QM}_{R_{10^{-20}}} \neq \emptyset$, $e^{QM}_{R_{10}} \neq \emptyset$, and $e^{QM}_{R_{10^6}} = \emptyset$. The domains of empirical adequacy of these theories thus differ.

The same theory may assign the same empirical content to different space-time regions. Consider the behavior of two boxes of gas, located in different rooms, for five minutes (say, space-time regions $R_1$ and $R_2$). The set of all history-descriptions $H^C$ of Classical Thermodynamics contains descriptions of the type ``there is a box of gas whose pressure, temperature, and volume is $(P_0, T_0, V_0)$ at time $t_0$ and is $(P_1, T_1, V_1)$ at time $t_1$" with some restrictions on the relationship between these values, such as $P_0 V_0 / T_0 = P_1 V_1 / T_1$. If both boxes of gas are in thermal equilibrium and happen to have the same pressure, temperature, and volume, then exactly the same history-descriptions in $H^C$ are empirically adequate to both rooms, and hence $e^{C}_{R_1} = e^{C}_{R_2}$. (Note that the empirical contents of the same space-time regions $R_1$ and $R_2$ differ according to Statistical Mechanics, since the particles making up the gas in the two boxes presumably move in different ways; thus $e^{SM}_{R_1} \neq e^{SM}_{R_2}$.)

As is clear from the examples, history-descriptions are relative to theories, and history-descriptions of different theories are often suggestive of different ontologies. History-descriptions of Newtonian physics suggest the existence of forces, but not of epicycles; vice versa for Ptolemaic physics. History-descriptions of Classical Thermodynamics suggest the existence of pressure and volume, but lack reference to particle positions and velocities of Statistical Mechanics. The difference between history-descriptions of different theories becomes even more pronounced with the special sciences: biology refers to organisms, macroeconomics to inflation, folk mental theories to pain, philosophical psychology to qualia, moral psychology to conscience, and so on.

\subsection{Empirical structure, empirical theory}\label{subsection_empiricalstructure}

Motivated by the previous discussion, I define the {\em empirical structure $e^T$ of theory $T$} as the partition composed of the sets of all space-time regions which have the same empirical content according to $T$; formulaically: $e^T ~ \doteq ~ \{ \{ R' \subseteq M ~|~ e^T_{R'} = e^T_R \} ~|~ \forall R \subseteq M  \}$.\footnote{
Thus extensionally the empirical structure of a theory only contains sets of space-time regions. My theory supervenience theses (see later) are expressible by making reference solely to empirical structure. Hence, while the above-given definition of empirical structure refers to empirical contents, which, as {\em sets} of history-descriptions, would necessitate a history-description to be a formal object, this latter assumption could be relaxed by defining empirical structure directly as consisting of sets of those space-time regions to which exactly the same history-descriptions are empirically adequate to (in the previously defined sense). I chose to keep the intermediary notion of empirical content for ease of exposition and to be able to develop a contrast between formal vs. empirical content and structure. See below, and cf. footnote \ref{footnote_firstperson}.
}

I understand the term {\em empirical theories} to refer both to theories of physics and to the {\em special sciences}, the latter of which I construe broadly to include all scientific theories other than physics, as well as any theory that has non-empty empirical content in the sense described above (e.g. folk theories invoking mental states).\footnote{\label{footnote_firstperson}
I primarily intend to understand the term empirical theory to refer to {\em third person view} theories. Thus, when I talk of mental theories as empirical theories, my primary focus is on theories akin to current theories of psychology or cognitive science: third person view descriptions of mental phenomena. 

However, the generality of my notion of a history-description and relativizing empirical adequacy to circumstances (to space-time regions) also opens up my account to be applicable to {\em first person view} mental or perceptual experiences. In this application my notion of empirical adequacy expresses a so-called accuracy condition and my notion of empirical content is a way to characterize what has been called the ``content of an experience" in the literature \citep[see][]{Siegel2024}. My notion of empirical structure would then characterize all those circumstances in the actual world in which a person would have the same contents of experiences. Understood this way, the thesis of empirical structure physicalism (see later) would connect the first person view with a third person view: persons can only have different contents of experiences in circumstances among which physics is also able to differentiate. (What allows connecting the first and the third person view is that in this application empirical adequacy would express an accuracy condition which connects first person view experiences with measurement reports that can be obtained from a third person view.)

Claims of this essay could also extend to theories involving moral and aesthetic values to the extent we can think of moral and aesthetic values as sort of things that physically embodied agents can hold and act upon as beliefs. The discussion could also be extended to theories involving abstract and transcendent entities, but I omit the details here with the note that \citet{Szabo2017,Szabo2021} gives a good starting point for their physicalist incorporation.
}
I assume that every empirical theory (can be reconstructed as to) have two types of content: formal and empirical. I take the {\em formal content} to be the set of all history-descriptions of the theory. The formal content of a theory may have certain structural properties; by choosing a certain type of structural property we may say that the formal content of a theory is a representation of its {\em formal structure}. Two theories are then {\em formally identical} if they have the same formal structure (for a number of different formal identity-criterions arising from different selections of what a formal structure is see \citealt[274]{Halvorson2019}). My notion of empirical structure gives an analogous criterion for identifying empirical theories whose empirical contents would differ as sets: it captures the idea that it is their empirical distinguishing ability which empirically differentiates theories. I thus regard two theories to be {\em empirically identical} if they have the same empirical structure. Two empirical theories can then be regarded identical if they are both formally and empirically identical.\footnote{
Identity-criterions of formal and of empirical structures are {\em prima facie} independent. Although in this paper I focus on relationships of empirical structure, this paper does not ascribe to the radical empiricist thesis that two theories are identical {\em tout court} if they have the same empirical structure (which would correspond to the Zenonian identification of different formal contents in the sense of \citealt[274]{Halvorson2019}). 
}

\subsection{Theory supervenience}\label{subsection_theorysupervenience}

I now introduce theory supervenience as a relation between empirical structures. It is easy to see that the following three conditions are equivalent:
\begin{itemize}
	\item[(i)] the empirical contents of space-time regions according to theory $T$ supervene on the empirical contents of space-time regions according to theory $T'$, that is, $\forall R_1, R_2 \subseteq M ~ : ~ e^T_{R_1} \neq  e^T_{R_2}  \rightarrow e^{T'}_{R_1} \neq e^{T'}_{R_2}$;
	\item[(ii)] there is no pair of space-time regions whose empirical contents are different according to $T$ but are the same according to $T'$, that is, $\nexists R_1, R_2 \subseteq M ~:~ e^{T}_{R_1} \neq e^{T}_{R_2} ~ \& ~ e^{T'}_{R_1} = e^{T'}_{R_2}$;
	\item[(iii)] the empirical structure of $T'$ refines the empirical structure of $T$ (in the usual sense of how one partition refines another partition), that is, $\forall {\cal R'} \in e^{T'} ~ \exists {\cal R} \in e^{T} : {\cal R'} \subseteq  {\cal R}$.
\end{itemize}
Condition (i) may be shortened as {\em theory $T$ supervenes on theory $T'$}, (ii) as {\em $T'$ renders $T$ empirically dispensable} (see discussion in Sect. \ref{section_mainarguments}), and (iii) as {\em $T'$ refines the empirical structure of $T$}; I will use these expressions interchangeably, and denote them with $T \precapprox T'$. If both $T \precapprox T'$ and $T' \precapprox T$, in notation if $T \approx T'$, then I say that $T$ and $T'$ {\em have the same empirical structure}. This terminology is apt since $T \approx T'$ if and only if $e^T = e^{T'}$. 

When $T \precapprox T'$ but not $T' \precapprox T$, in notation $T \prec T'$, I analogously say that (i) {\em $T$ strongly supervenes on $T'$}, or (ii)  {\em $T'$ renders $T$ both empirically dispensable and non-fundamental}, or (iii) {\em $T'$ strongly refines the empirical structure of $T$}. The second terminology comes from the following observation: {\em  $T'$ renders $T$ non-fundamental} if there is a pair of space-time regions $R_1$ and $R_2$ such that $T$ is not able to empirically distinguish between $R_1$ and $R_2$, but $T'$ is able to, that is, if $~\exists R_1, R_2 \subseteq M ~ : ~ e^{T}_{R_1} = e^{T}_{R_2} ~ \& ~ e^{T'}_{R_1} \neq e^{T'}_{R_2}$).

Clearly, theory supervenience and strong theory supervenience are reflexive and transitive relations. 

For any pair of empirical theories one can ask whether one theory supervenes on the other. It is instructive to ask this question both for diachronic and for synchronic examples, such as former vs. later theories of physics (which intended to have the same domain of empirical adequacy), and for `higher' vs. `lower' level (macroscopic vs. microscopic) theories.

{\em Example for diachronic theory supervenience}: Newtonian gravitational theory supervenes on General Relativity. To see this, as before, let $R_{10}$ be a space-time region containing the motion of planets of our Solar system for the span of ten years, and let $e^{N}_{R_{10}}$ and $e^{GR}_{R_{10}}$ be the empirical contents of $R_{10}$ according Newtonian gravitational theory and General Relativity, respectively. Let us now consider another space-time region $R'_{10}$ in our universe (for instance, our own Solar system again, but in the past or future relative to $R_{10}$), and suppose that the empirical contents of $R_{10}$ and $R'_{10}$ differ according to Newtonian gravitational theory: $e^{N}_{R_{10}} \neq e^{N}_{R'_{10}}$. Question: is it also the case that the empirical contents of $R_{10}$ and $R'_{10}$ also differ according to General Relativity, that is, $e^{GR}_{R_{10}} \neq e^{GR}_{R'_{10}}$? I claim this is the case: $e^{N}_{R_{10}} \neq e^{N}_{R'_{10}}$ but $e^{GR}_{R_{10}} = e^{GR}_{R'_{10}}$ would entail that Newtonian gravitational theory can differentiate the two space-time regions in an empirically successful way but General Relativity does not have this ability. If General Relativity indeed inherits all empirical successes of Newtonian gravitational theory, as is widely believed, then such a case cannot arise for any pair of space-time regions. Hence Newtonian gravitational theory supervenes on General Relativity. (The converse does not hold, since there are pairs of large space-time regions, e.g. the aforementioned $R_{10^6}$ and another similarly sized $R'_{10^6}$, for which $e^{GR}_{R_{10^6}} \neq e^{GR}_{R'_{10^6}}$, but $e^{N}_{R_{10^6}} = e^{N}_{R'_{10^6}} = \emptyset$. Hence Newtonian gravitational theory also strongly supervenes on General Relativity.)

{\em Example for synchronic theory supervenience}: Classical Thermodynamics strongly supervenes on Statistical Mechanics. We already illustrated this case in Sect. \ref{subsection_empiricalcontentexample}, since it is not possible that Statistical Mechanics empirically adequately describes two boxes of gas as having identical microscopic properties yet the boxes of gas are empirically adequately described by Classical Thermodynamics as having different temperature, pressure, or volume. 

By {\em physics at a certain time} I understand the collection of all theories of physics which are generally accepted at that time. In this article I will make the claim that earlier theories of physics supervene on later theories of physics. To make this claim precise, I define that {\em theories $\{ T_i \}_{i \in I}$ supervene on theories $\{ T'_j \}_{j \in J}$} (alternatively: {\em $\{ T'_j \}_{j \in J}$ renders $\{ T_i \}_{i \in I}$ empirically dispensable}, or {\em $\{ T'_j \}_{j \in J}$ refines the empirical structure of $\{ T_i \}_{i \in I}$}) if, whenever the empirical contents of a pair of space-time regions differ according to any of the theories in the first collection, then the empirical contents of the same pair of space-time regions also differ according to at least one theory in the second collection. That is, $\{ T_i \}_{i \in I} \precapprox \{ T'_j \}_{j \in J}$ if $\forall R_1, R_2 \subseteq M ~ : \left( ~\exists i \in I:  e^{T_i}_{R_1} \neq  e^{T_i}_{R_2}  \rightarrow \exists j\in J: e^{T'_j}_{R_1} \neq e^{T'_j}_{R_2} \right)$. (Mutatis mutandis for strong supervenience.) 

Finally, when physics at time $t$ renders a theory $T$ empirically dispensable I simply say that {\em $T$ is empirically dispensable at time $t$}, and when physics at time $t$ renders $T$ non-fundamental I simply say that {\em $T$ is non-fundamental at time $t$}.

\subsection{Empirical structure realism, empirical structure physicalism}\label{subsection_theorysuperveniencephysicalism}

The two examples of the previous subsection suggest two groups of theses regarding theory supervenience. The first group asserts a type of historical-sociological continuity between earlier and later physics. They can be expressed in equivalent ways (see Sect. \ref{subsection_theorysupervenience}):
\begin{itemize}
	\item[] {\em (R$_p$) Past empirical structure realism}: up until the present, \\
			$~~~~~~~~$ -- later physics refined the empirical structure of earlier physics; or \\
			$~~~~~~~~$ -- earlier physics supervened on later physics; or \\
			$~~~~~~~~$ -- later physics rendered earlier physics empirically dispensable.
	\item[] {\em (R$_f$) Future empirical structure realism}: future physics will refine the empirical structure of current physics.
		\\
		Combining the two theses yields:
	\item[] {\em (R) Empirical structure realism}: later physics refines the empirical structure of earlier physics.
\end{itemize}
Thesis (R) can be thought of as a counterpart to structural realism and to scientific realism: while structural realism asserts a retention of a certain structure in the formal content of theories of physics across theory change \citep{Ladyman2023}, and while scientific realism asserts a retention of certain entities suggested by the formal content of theories of physics across theory change, empirical structure realism asserts a retention of empirical structure of physics across theory change. Thesis (R) thus expresses a cumulative account of theory change (cf. Kuhnian appraisals).

The second group of theses asserts a type of hierarchy between `higher' and `lower' level theories:
\begin{itemize}
	\item[] {\em (P$_c$) Currentist empirical structure physicalism}:  current special sciences \\
			$~~~~~~~~$ -- are currently empirically dispensable;\footnote{
(P$_c$) says in particular that current physics renders current mental theories empirically dispensable. Some authors, most relevantly \citep[10]{Firt-Hemmo-Shenker2022}, appear to suggest this cannot be taken for granted for the reason that ``[in] von Neumann's (1932) standard formulation of quantum mechanics, which is our best contemporary fundamental framework of quantum field theory, the {\em mental} is an indispensable part of the physical theory [...]". An anonymous referee of this paper raised a similar point.

I disagree with this characterization. No standard physics textbook explicitly presents or understands the mental either as an indispensable part of quantum mechanics or as belonging to its standard formulation. Despite the authors' suggestion, this includes the referenced book by \citet{vonNeumann1932}, which does not contain a single mention of the words ``mind", ``mental" or ``consciousness". Without going into a detailed exegesis of von Neumann's presentation of the so-called Heisenberg cut (a gap in the understanding of quantum mechanical time development that some may hope to fill in by attaching significance to the consciousness of an observer), I remark that even the sole article referenced by \citep[10]{Firt-Hemmo-Shenker2022} as an interpretation which ``propose[s] the conjecture that the collapse of the quantum state is triggered by the mind" agrees that von Neumann ``does not clearly identify measurement with conscious perception" and that his remarks ``suggests neutrality on whether the collapse process is triggered by measuring devices or by conscious observers" \citep[4]{Chalmers-McQueen2022}. The same article also explicitly admits that the idea that consciousness collapses the quantum wave function ``is now widely dismissed" \citep[1]{Chalmers-McQueen2022}.

Beyond standard physics textbook presentations it is indeed the case that serious physicists, most notably \citep{Wigner1962}, explicitly pondered the possibility that consciousness plays a role in the collapse of the wave function. However, these speculations were always carefully qualified as such by the physicists who proposed them. For instance, Wigner --- whose oft-cited article was written for an anthology aptly titled {\em The Scientist Speculates: An Anthology of Partly-baked Ideas} --- wrote he ``is prepared to admit that [whether the relation of mind to body will enter the realm of scientific inquiry] is an open question" \citep[284]{Wigner1962}, that ``[it] may be premature to believe that the present philosophy of quantum mechanics will remain a permanent feature" (285), and that he ``would not be greatly surprised if [his] article shared the fate of those of his predecessors" which ``were either found to be wrong or unprovable, hence, in the long run, uninteresting" (298).

More pertinently, there is a claim which is widely accepted in the foundations of physics, namely that there are multiple interpretations of quantum mechanics which do not have a Heisenberg cut, for instance, Bohmian mechanics. Empirical indistinguishability of these interpretations entails that any mental-permissive interpretation supervenes on Bohmian mechanics. Thus we may safely assume that if any current special science (such as a mental theory) supervenes on any version of ordinary quantum mechanics, then it also supervenes on Bohmian mechanics, which does not invoke mental descriptions. Hence mental descriptions are not indispensable in quantum mechanics. 
}
			or	\\
			$~~~~~~~~$ -- supervene on current physics; or \\
			$~~~~~~~~$ -- have an empirical structure which is refined by current physics.
	\item[] {\em (P$_f$) Futurist empirical structure physicalism}: the current special sciences are empirically dispensable in the future.
	\\
	Combining the two theses yields:
	\item[] {\em (P) Empirical structure physicalism}: the current special sciences are empirically dispensable, both currently and in the future.\
\end{itemize}
Thesis (P) can be thought of as a counterpart to linguistic and to theory-oriented metaphysical physicalism: while linguistic physicalism asserts a dependency relationship between the formal content of mental theories and the formal content of theories of physics, and while theory-oriented metaphysical physicalism asserts a dependency relationship between the entities suggested by the formal content of mental theories and the entities suggested by the formal content of theories of physics, empirical structure physicalism asserts a dependency relationship between the empirical structure of mental theories and the empirical structure of physics.

In the interest of brevity I only formulate here the supervenience versions of these theses; of course, the strong supervenience versions may be of more interest. The strong supervenience version of (P) is not a great leap for someone already accepting the supervenience version (P). After all, a difference of a lone ammonium molecule between two rooms \citep[cf.][]{Stoljar2021} is not likely to change any of the current special sciences' empirical contents, but it does change the empirical content of molecular-level current physics.

Thesis (P) does not state that terms employed by mental history-descriptions cannot become an integral part of a future physics. Classical Thermodynamics is also currently empirically dispensable (and thus, if future empirical structure realism holds, it remains empirically dispensable in the future as well). Yet, we still regard notions of pressure and temperature as belonging to current physics, since Classical Thermodynamics belongs to the collection of currently accepted theories of physics. Thus, even if thesis (P) were true, it may still be the case that mental terms will find a place similar to temperature or pressure in a future physics --- however, they will remain  empirically dispensable.

\section{Empirical structure physicalism and Hempel's dilemma}\label{section_mainarguments}

After much preparation, I can now investigate the status and relationships of the theory supervenience theses introduced in Sect. \ref{subsection_theorysuperveniencephysicalism}.

Currentist empirical structure physicalism (P$_c$) is not rendered false by incompleteness of current physics (by premise (P3) of Hempel's dilemma). Ontological posits of theories are suggested by their history-descriptions, but (P$_c$) is neither a thesis about sameness of (aspects of) history-descriptions assigned to space-time regions, nor a thesis about sameness of empirical contents assigned to space-time regions. Hence, (P$_c$) is not a metaphysical thesis. Theory supervenience of the special sciences on current physics does not require that they share {\em any} of their history-descriptions; it only requires every {\em difference} of empirical contents of a {\em pair} of space-time regions according to the special sciences to be registered as a {\em difference} of empirical contents of the {\em same pair} of space-time regions according to current physics. In short, empirical structure physicalism claims that physics refines the empirical {\em structure} of the special sciences, but an empirical structure does not reference any history-description: it is only composed of sets of space-time regions. (P$_c$) is thus independent of whether the special sciences and physics have the same ontological posits suggested by their history-descriptions, or have the same history-descriptions, or have the same sets of history-descriptions characterizing any {\em single} space-time region. 

In sum, currentist empirical structure physicalism (P$_c$) does not contradict premise (P3) of Hempel's dilemma as formulated in Sect. \ref{section_hempelsdilemma}. 

One could object that currentist empirical structure physicalism (P$_c$) is still subject to a {\em reformulated} threat of future changes in physics, tailored to currentist empirical structure physicalism. If futurist empirical structure physicalism (P$_f$) were false, then it may be the case that current mental theories do not supervene on future physics. If current mental theories did not supervene on future physics, then there would be a pair of space-time regions $R_1$ and $R_2$ and a history-description $h$ of current mental theories such that $h$ is empirically adequate to $R_1$ but not to $R_2$, yet future physics is unable to empirically identify any difference between $R_1$ and $R_2$. In this case, current mental theories would become empirically {\em in}dispensable. Such an $h$ would also presumably suggest the existence of a mental entity, such as the presence of pain or of a qualia, and the mental theory would only assign this entity to $R_1$, but not to $R_2$. Thus, in this case, our most comprehensive future understanding of what exists in our world would require an appeal to mental entities which are empirically indispensable in the future. 

{\em Prima facie} arguing for futurist empirical structure physicalism (P$_f$) may seem even more hopeless than arguing for currentist empirical structure physicalism (P$_c$). As premise (P4) of Hempel's dilemma points out, future theories of physics are currently unknown. Thus, if we take (P$_f$) to assert an {\em a posteriori} thesis, it is possible that future physics changes in a way to render (P$_f$) false. On the other hand, positing (P$_f$) as an  {\em a priori} thesis would clearly be question begging.

This is where the first key observation of this article steps in. The first key observation is that futurist empirical structure physicalism (P$_f$) is a simple deductive consequence of currentist empirical structure physicalism (P$_c$) and future empirical structure realism (R$_f$). This is so since if the current special sciences supervene on current physics (P$_c$), and if current physics supervenes on future physics (R$_f$), then it follows from transitivity of theory supervenience that the current special sciences also supervene on future physics (P$_f$). (Mutatis mutandis for strong supervenience.)

Hence, for someone who accepts currentist empirical structure physicalism (P$_c$), it is sufficient to defend future empirical structure realism (R$_f$) to guard against the threat of future changes in physics, as well as to show that mental theories remain empirically dispensable in the future (and also non-fundamental in the case of the strong supervenience version of (P)).

{\em Prima facie} we just exchanged one difficulty for another: arguing for future empirical structure realism (R$_f$) may also seem to be hopeless. Again, taking (R$_f$) to be an {\em a posteriori} thesis, it is possible that future physics changes in a way to render (R$_f$) false. On the other hand, positing (R$_f$) as an {\em a priori} thesis would clearly be question begging.

This is where the article's second key observation comes into play. The second key observation is that, even though it is {\em logically possible} that future empirical structure realism (R$_f$) is false, (R$_f$) is a strong meta-inductive consequence of the {\em a posteriori}, empirically verifiable, and plausible historical claim that up until the present, later physics refined the empirical structure of earlier physics (R$_p$). I postpone the discussion to Sect. \ref{section_optimisticmetainduction}. 

Let me briefly summarize where we have arrived. So far, I argued that {\em if} we accepted currentist empirical structure physicalism (P$_c$) and past empirical structure realism (R$_p$), then futurist empirical structure physicalism (P$_f$) becomes strongly inductively supported, and the reformulated threat from future changes of physics can also be avoided. Hence, currentist and futurist empirical structure physicalism can be combined, and we can conclude that empirical structure physicalism (P) avoids both Hempel's dilemma as formulated in Sect. \ref{section_hempelsdilemma}, as well as how Hempel's dilemma can be reformulated as a threat against empirical structure physicalism.\footnote{Thus, although empirical structure physicalism is a claim about the relationship between the empirical structure of the special sciences and the empirical structure of physics, and this relationship is a changeable account of experience in the sense of {\em logical} possibility, it may not be a changeable account of experience in the sense of {\em empirical} possibility \citep[see][]{Norton2022,Gyenis2025}; \citep[cf.][]{Firt-Hemmo-Shenker2022}.}

Thus, (P), (P2)-(P4) form a consistent set of propositions. I postpone the discussion whether (P), (P2)-(P4) indeed respects all of our defensible epistemic attractions behind the original inconsistent set of propositions (P1)-(P4), and hence whether (P) not only {\em avoids}, but also {\em solves} Hempel's dilemma, until Sect. \ref{section_comparison}.

Finally, although in this article I do not intend to give conclusive arguments for past empirical structure realism (R$_p$) or for currentist empirical structure physicalism (P$_c$), I argue for their plausibility in Sections \ref{section_optimisticmetainduction} and \ref{section_comparison}, respectively.

\section{Empirical structure realism and an optimistic meta-induction}\label{section_optimisticmetainduction}

According to future empirical structure realism (R$_f$), whenever current physics is able to differentiate between the happenings of a pair of space-time regions in an empirically successful way, future physics --- regardless of the new entities, properties, processes, or empirical methods it may invoke --- will also retain the ability to recognize such differences in an empirically successful way. In other words, future empirical structure realism only fails if future physicists decided that a current or future theory of physics, which is empirically successful in differentiating empirical content, needs to be completely discarded, even though none of the other accepted theories of physics have this differentiation ability. It seems unlikely that the community of physicists, in the absence of serious and disruptive societal pressure, would ever completely discard a theory if doing so resulted in a loss of success in empirical content differentiation. Although improving the ability to successfully draw contrasts between empirically different scenarios may not be the central epistemic goal of all sciences, it is part of the ethos gluing together the community of physicists.

One could object that so-called {\em Kuhn losses} --- empirical or theoretical successes of a replaced theory of physics that did not get carried over to the replacing theory of physics --- put this intuition in doubt. However, the existence of Kuhn loss in historical cases is heavily debated in the literature \citep[see i.e.,][]{Veronen1992}; even if a Kuhn loss existed, it is unclear whether it would concern empirical success in the narrow sense needed by thesis (R). All putative examples of Kuhn loss I am aware of either entail a loss in explanatory power without a loss in predictive capability, or are not genuine examples of loss in the relevant sense, as they concern equations (e.g., the Poiseuille equation) that were not discarded upon the acceptance of a new theory \citep[see][\pp 111-114]{Votsis2011}, and hence still belonged to the physics of the time.

Future empirical structure realism can also be straightforwardly argued for on the basis of an {\em optimistic meta-induction} from historic empirical structure realism: simply put, thesis (R$_p$), via meta-inductive inference, entails thesis (R$_f$).

The optimistic meta-induction relies on the historical premise (R$_p$). Length limitations prevent me from defending the historical premise here, but with a few qualifying remarks I intend to illustrate its plausibility. 

Some rational reconstruction is needed to defend the historical premise, but its extent does not seem to put the generalization in serious doubt. One potential issue is the ability of competent theory users to recognize whether different history-descriptions have the same empirical content.\footnote{
The {\em empirical content $e^T_h$ of a history-description $h$ according to $T$} is the set of all history-descriptions that are empirically adequate to the same space-time regions as $h$; formulaically: $e^T_h ~ \doteq ~ \{ h' \in H^T ~|~ \forall R \subseteq M ~ : ~ EA^T(h',R) \leftrightarrow EA^T(h,R) \}$.
}
 A history-description of Ptolemaic astronomy that utilizes an equant has the same empirical content as a history-description that has an additional epicycle instead of an equant \citep[see][]{Rabin2019}: with such a replacement we get two, formally different history-descriptions that suggest different ontologies but have the same empirical content. The same problem arises with any theory which has gauge freedom, that is, with most modern theories of physics \citep[for an illustrative introduction to the philosophy of physics literature on gauge freedom, see][]{Norton2019}. 

In reply, while practitioners of a given theory of physics may not have realized the sameness of empirical content for a period of time, doing so does not conceptually require the perspective of a new, superseding theory, only the normal practice of getting a better understanding of the given theory, the presence of superfluous mathematical structure and mathematical equivalences in its history-descriptions, and their representational relationship with the happenings they aim to represent. For instance, it did not require the appearance of General Relativity to understand, as a consequence of the Newton-Leibniz debate, that while the Newtonian view that absolutizes both velocity and acceleration leads to a multiplicity of history-descriptions that share the same empirical content, the Leibnizian view that relativizes both velocity and acceleration leads to a deficit in empirical adequacy, and so the proper formulation of classical mechanics is one that relativizes velocity but absolutizes acceleration, for which the appropriate mathematical representational structure can also be given \citep[see][]{Earman1989}. While the question of whether different history-descriptions of a theory correspond to distinct {\em ontological} possibilities may not be resolved for a long time (given that such resolution typically requires invoking certain philosophical principles, such as the identity of indiscernibles \citep{Forrest2010}, and physicists frequently treat such principles with suspicion), a mature theory's practitioners usually agree on whether different history-descriptions differ in their {\em empirical} content. But that is all my historical premise (R$_p$) needs.

Similarly, some rational reconstruction may also be needed in order to distinguish unjustified belief in empirical adequacy and justified belief in empirical adequacy: for instance, early practitioners of a theory may have been convinced that certain history-descriptions have different empirical content, even though this is not the case. A potential example is Blondlot's theory of N-rays \citep[see][]{Nye1980}, a form of radiation hypothesized at the beginning of the 20th century. Blondlot was convinced that he had developed a reliable empirical practice to tell when N-ray radiation was present in certain laboratory scenarios. Blondlot's empirical practice, however, was soon shown to be unreliable by a visiting physicist, who, in essence, pointed out that basic, common-sense practices guarding against cognitive bias were not observed, and Blondlot's ability to differentiate disappears as soon as these common sense practices are introduced. 

The codification of such basic, common-sense practices guarding against cognitive bias as elementary statistical procedures has matured only in the past century, and hence projecting such practices back as requirements for proper judgments of empirical adequacy in historical examples is historically anachronistic. However, these elementary statistical procedures are independent of the entities, properties, laws, processes, mechanisms etc. postulated by the scrutinized theories: they are merely tools to filter out cognitive biases and spurious correlations. Hence even if a counterexample against my historical premise were found in the history of physics, as long as it is a counterexample only due to the presence of an unjustified belief in empirical adequacy, but this unjustified belief could have been guarded against by employing common sense statistical practices, my historical premise is in no real danger of falsification.

I close this section with contrasting my optimistic meta-induction with the well-known pessimistic meta-induction. The two arguments share the same form: that of an enumerative induction over past theories of physics. However, their inductive base is different: while the pessimistic meta-induction focuses on survival of entities, properties, and references of certain terms, the optimistic meta-induction focuses on retention of empirical structure. Since an empirical structure can be refined during a theory change even if entities, properties, and references of certain terms do not survive, there is no conflict between my optimistic and the pessimistic meta-induction: both could hold at the same time. For my purposes here this is enough.\footnote{Even though the two arguments could coexist, the pessimistic meta-induction is arguably much weaker as an inductive argument than my optimistic meta-induction. This difference in inductive strength can be articulated in terms of the material theory of induction \citep{Norton2003b}. \citet{Shech2019} argues that the pessimistic meta-induction is (at best) a weak inductive argument since it lacks a licensing material fact. For the optimistic meta-induction, material facts that could license the induction are evidently given by the sociology of physics: as mentioned before, it is very unlikely that the community of physicists, in the absence of some serious disturbing societal pressure, will ever completely discard and replace a theory of physics with another if this resulted in a loss of empirical success in differentiation. However, details should be pursued elsewhere.}

\section{Physicalisms, dualisms, and multiple realization}\label{section_comparison}

In Sect. \ref{section_mainarguments} I argued that empirical structure physicalism (P) avoids Hempel's dilemma in the sense that premises (P), (P2)-(P4) form a consistent set of propositions. Here I ask to what extent (P), (P2)-(P4) respect our defensible epistemic attractions behind the original inconsistent set of propositions (P1)-(P4).

\subsection{Physicalisms}\label{subsection_physicalisms}

In the ever-growing literature on physicalism we find many subtly different physicalism theses. This cornucopia has recently prompted philosophers to assemble properties which they regard as `desirable' for a thesis of physicalism or for the notion of `physical' (property). Thus, one way to assess whether empirical structure physicalism indeed deserves to be labeled as a type of physicalism is by checking how many of these `desirable' properties it has.

Here is such an assembled list of desirable properties (for items 2-5 cf. \citet[\p 82]{Marton2019} and for items 5-13 cf. \citet[\pp 46-50]{Horzer2020}), suitably rephrased for our discussion of empirical structure physicalism:
\begin{itemize}
	\item[1.] It expresses a type of reductive relationship between the special sciences and physics.
	\item[2.] Its scope is determined by the theories we include in the set of special sciences.
	\item[3.] Its content is determined by the empirical contents that theories provide.
	\item[4.] It respects our everyday intuition about the type of phenomena physics is concerned with. 
	\item[5.] It is not trivially or obviously false.
	\item[6.] It is contingent, {\em a posteriori}, and not trivially true.
	\item[7.] It is not defined as a sort of heuristic or oath towards science (it is truth-apt).
	\item[8.] Progress of physics does not render it automatically true or false.
	\item[9.] It does not rule out a proper distinction of physicalism from dualism and idealism (and other traditional rivals of physicalism).
	\item[10.] It does not incorporate special sciences into physics in a trivial way.
	\item[11.] It could have different truth value if physics were different.
	\item[12.] It does not presume an {\em a priori} mereological hierarchy. 
	\item[13.] It does not exclude {\em a priori} the possibility that physics could contain dispositional properties.
\end{itemize}
In addition to these general properties, the following particular properties can also be formulated based on challenges that prior articulations of currentist physicalism needed to face \citep[see][\pp 84-88]{Marton2019}:
\begin{itemize}
	\item[14.] It does not assume that physics of ordinary matter is essentially complete \citep[cf.][]{Smart1978}.
	\item[15.] It does not assume that we can regard the presently known neural structures as sufficient bases for understanding mental phenomena \citep[cf.][]{Lycan2003}.
	\item[16.] It does not exclude quantum mechanical phenomena from being relevant for an explanation of consciousness.
	\item[17.] It does not exclude the possibility that certain entities influence the central nervous system differently than other physical systems.
	\item[18.] It is compatible with multiple realization, and with the possibility of quantum computers realizing mental states.
	\item[19.] It does not rely on assumptions about the probability of truth of current or future theories of physics.
	\item[20.] It does not rely on the assumption that physicalism is a general empirical hypothesis.
\end{itemize}
After an extensive discussion of various views on what physical properties amount to, in relation to counterparts of criteria 5-13 above, \citet[\p 107]{Horzer2020} concludes that ``[T]he views that are most prominently discussed in the literature all turn out to fail with regard to one or the other criterion that a proper account of the physical needs to satisfy." That is, all prominently discussed views fail to respect at least one of the above-listed desirable properties (or their suitable reformulations).

It only takes minimal reflection to see that empirical structure physicalism satisfies at least 18 out of the above-listed 20 requirements. The only two questionable items are item 9 (relationship to dualism) and item 18 (compatibility with multiple realization).

\subsection{Dualisms}\label{subsection_dualisms}

Suppose the following happens in an isolated room of a research facility during the span of ten minutes (space-time region $R_1$): initially, there is a lady sitting next to a table with a pen and pencil. At the two-minute mark, a question is asked of her through a built-in loudspeaker. She answers the question. At the five-minute mark, a loud beep is heard. At the eighth-minute mark, another question is asked of her through the loudspeaker. She answers this second question. 

The research facility also has another room (different space-time region $R_2$). Assume that the two rooms have the same empirical content both according to current physics and according to current mental theories during the first two minutes. When we say that the empirical content of the rooms $R_1$ and $R_2$ is the same during its first two minutes according to current physics, this entails that all particles and fields in the first two minute subregions of $R_1$ and of $R_2$ (for the entire first two minutes in the two rooms) have, as far as any measurement accepted by any of the current theories of physics is concerned, the same distribution. This includes the particles and fields making up the bodies and brains of the lady in room $R_1$ and the lady in room $R_2$, how the sounds they make move the air, and so on. When we say that the empirical content of both rooms is the same during its first two minutes according to current mental theories, this similarly means that, as far as any measurements accepted by current mental theories is concerned, there is no difference between the first two minute subregions of $R_1$ and $R_2$. 

For instance, {\em one} of the many measurement methods accepted by current psychology is interview questions asked in controlled environments. Such questions could be asked through the loudspeaker, and an empirical difference between the two rooms (according to mental theories) could be established based on the answers. Thus, for the two rooms to have the same empirical content according to mental theories it needs to be the case that if we (counterfactually) asked any question through the loudspeaker from the lady in room $R_1$ and the exact same question from the lady in room $R_2$ then we should receive identical answers. 

Suppose, furthermore, that in the first two minutes the ladies in both rooms know everything current physics says about the color red, but neither of them have ever experienced red as a color before (in analogy with the so-called knowledge argument, see \citealt{NidaRumelin-Conaill2019}). It is possible, through measurement methods accepted by current psychology, to confirm whether this is indeed an empirically adequate history-description of mental theories. For instance, we could ask any question about current physics through the loudspeaker at the two-minute mark and confirm whether the ladies indeed give the correct answers; similarly, we could also ask any question about the past experiences of the ladies and confirm that they have, indeed, never experienced red as a color. 

Finally, suppose that the two rooms have the same empirical content according to current physics during the remaining eight minutes as well. Does this entail that the empirical content of the two rooms is also same according to current mental theories for the remaining eight minutes? For instance, could the mental history-description ``the lady has the qualia red" be empirically adequate to the first room according to current mental theories during the three minutes elapsing between the beep at the five-minute mark and the second question of the loudspeaker at the eighth-minute mark, but not empirically adequate to the second room during the same three minutes?

According to empirical structure physicalism (P) the answer to this question is negative: if the current special sciences supervene on physics, then if the empirical contents of $R_1$ and $R_2$ are the same according to current physics, then their empirical contents must also be the same according to current mental theories.

One does not need to invoke metaphysical principles (such as the causal closure of the physical) to see why this negative answer is plausible. For instance, if the loudspeaker at the eighth-minute mark asked the question ``has your knowledge of red changed since you heard the loud beep" in both rooms, and the lady in the first room answered ``yes" while the lady in the second room answered ``no", then the sound waves of these different answers would move the air in the two rooms differently, and thus the two rooms could not have the same empirical content according to current physics. This is also the case for any other measurement methods, since no measurement method exists for current special sciences whose different outcomes would not manifest in a difference of empirical contents of current physics.\footnote{Of course, if after the loud beep someone opened the door of the first room and showed a red object to the lady sitting there, but no one showed a red object to the lady sitting in the second room, then it would be no surprise that the lady in the first room answered the question with a ``yes'', and the lady in the second room answered it with a ``no". But in this case the empirical contents of $R_1$ and $R_2$ would also differ according to current physics. Thus, such an example would not violate empirical structure physicalism (P).}

This is my short plausibility argument for empirical structure physicalism. Nevertheless, there is a clear philosophical thesis which can be properly distinguished from empirical structure physicalism, namely:

{\em (D) Empirical structure dualism}: current mental theories do not supervene on physics.

Thesis (D) can not only be properly distinguished from thesis (P), but it is also the case that if (P) is true then (D) must be false; conversely, if all current special sciences other than current mental theories supervene on physics but (P) is false then (D) must be true. Thus, to this extent, empirical structure physicalism also satisfies requirement 9 of the previous subsection.

An objection philosophers of mind would surely raise is that even if (P) were true, {\em metaphysical} dualism could still also be true. To see what this claim amounts to, here is a way one can formulate a (sufficient condition for a) metaphysical dualist thesis:

{\em (MD) Metaphysical dualism}: there is a pair of space-time regions $R_1$ and $R_2$ and a mental entity $m$ such that $m$ exists in $R_1$, $m$ does not exist in $R_2$, but exactly the same physical entities exist in $R_1$ and in $R_2$.

Suppose $R_1$ and $R_2$ are space-time regions appearing in the definition of (MD), that is, a mental entity $m$ exists in $R_1$, $m$ does not exist in $R_2$, but exactly the same physical entities exist in $R_1$ and $R_2$. Since exactly the same physical entities exist in $R_1$ and in $R_2$, the empirical contents of $R_1$ and $R_2$ must be the same according to current physics. Then it follows from empirical structure physicalism (P) that the empirical contents of $R_1$ and $R_2$ are the same according to every current empirical theories, including current mental theories. But then if $T$ is a current empirical theory, $h$ is a history-description of $T$ which suggests the existence of $m$, and $T$ asserts that $h$ is true of $R_1$ but not of $R_2$, then this theory does so even though $R_1$ and $R_2$ have the same empirical content according to the same theory $T$! Such a theory is not successful as an empirical theory describing the differences between $R_1$ and $R_2$.

Thus, although it is logically possible that both metaphysical dualism (MD) and empirical structure physicalism (P) are true, believing that they are both true runs counter to what we may call: 

{\em (O) Occam's Contrastive Razor}: one should (currently) assign low confidence to the proposition that an entity $m$ exists in space-time region $R_1$ but $m$ does not exist in space-time region $R_2$ if either (1) the existence of $m$ is not asserted by any current empirical theory, or if (2) for every current empirical theory which asserts that $m$ exists in $R_1$ but not in $R_2$, the empirical contents of $R_1$ and $R_2$ are the same according to the theory.

I assume Occam's Contrastive Razor is an epistemically attractive proposition which can be motivated independently of the context of Hempel's dilemma. 

Thus, although it is possible to believe that both metaphysical dualism (MD) and empirical structure physicalism (P) is true, according to Occam's Contrastive Razor (O), one should assign low confidence to this belief. But then the epistemic attraction behind holding both (MD) and (P) is not defensible. Combining this with the claim that empirical structure physicalism appears to respect all other epistemic attractions behind theory-oriented metaphysical physicalism (P1)-(P2) (at least as far as the 20 item list of the previous subsection is concerned), we arrive to the conclusion that propositions (P), (P2)-(P4) respect all of our {\em defensible} epistemic attractions behind the original inconsistent set of propositions (P1)-(P4).

Hence, if we understand the challenge of Hempel's dilemma as a challenge of finding a type of reductive relationship between the special sciences and physics which can replace theory-oriented metaphysical physicalism and avoid the dilemma, as I do in this paper, then empirical structure physicalism solves Hempel's dilemma (notwithstanding that this solution may not satisfy certain philosophers of mind who are only willing to understand physicalism and dualism as {\em metaphysical} doctrines, and thus who are only willing to understand Hempel's dilemma as a challenge against theory-oriented metaphysical physicalism).

All being said, empirical structure physicalism (P) is not a metaphysical thesis. The linguistic physicalisms of Hempel, Carnap, and Neurath are not metaphysical theses either, yet they still have been labeled theses of physicalism \citep[82--83]{Horzer2020}. I submit that empirical structure physicalism is also deserving of the label, but in the end I worry less about labeling and more about the fact that the thesis avoids Hempel's dilemma yet saves all (or at least, most) of our defensible epistemic attractions behind theory-oriented metaphysical physicalism as a statement of a reductive relationship between the special sciences and physics.\footnote{I do not claim that empirical structure physicalism expresses the most informative reductive relationship which could solve Hempel's dilemma in this sense. Clearly, we would arrive at a more informative reductive relationship if, in addition to refinement of empirical structure, we provided more details about how the formal structures of the special sciences relate to the formals structures of theories of physics (although any such addition risks running again into Hempel's dilemma). For developing such an account between the causal claims made by the special sciences and physics, \citet{Fazekas-Gyenis-Szabo-Kertesz2021} is a good starting point. 
}

\subsection{Multiple realization}\label{subsection_multiplerealization}

To develop a concrete example of multiple realization, let me start by disambiguating the term `computer' as both a formal and an empirical concept. As a formal concept the term computer may refer to a Turing machine, to a finite Turing machine, or to some other similar, purely mathematical construction; the term as such belongs to the formal content of a theory. As an empirical concept a computer refers to some description of a part of our world given by one of our many theories of physics; these theories of physics distinguish parts of our world which are empirically adequately described as containing a physically embodied computer from those which are not. Thus, a computer as an empirical concept may refer to ordinary computers such as my laptop --- which fall into the domain of empirical adequacy of many different current theories of physics ---, or, hypothetically, to general relativistic computers (whose moving parts would be made up of planets and which occupied very large space-time regions, such as $R_{10^6}$ and $R_{10^6}'$ of Sect. \ref{subsection_theorysupervenience}), quantum-sized computers (whose moving parts would be made up of tiny configurations of atoms and which occupied very small space-time regions, such as $R_{10^{-20}}$ and $R_{10^{-20}}'$ of Sect. \ref{subsection_theorysupervenience}), or other types of exotic computers which only fall into the domain of empirical adequacy of one of our currently accepted theories of physics. What connects all of these theories of physics as empirical theories is that their formal structures share the same formal computer concept; but since these theories are also concerned with the question whether computer as a description is empirically adequate to a certain part of our world, they are empirical theories nevertheless. 

Thus current computer science, understood as an empirical rather than a purely formal theory, is a current special science which supervenes on current physics according to empirical structure physicalism (P). However, thesis (P) does not state that the current special sciences supervene on every currently accepted theory of physics, only that they supervene on at least one of them. For instance, our hypothetical general relativistic computers only supervene on General Relativity, while our hypothetical quantum-sized computers only supervene on Quantum Mechanics. 

We do not currently have a fundamental theory of physics on which all other current theories of physics supervene, and it is an open question whether we will ever have such a theory. In particular, we do not have a single current theory of physics which could empirically adequately describe both the difference between the large space-time regions $R_{10^6}$ and $R_{10^6}'$ and between the small space-time regions $R_{10^{-20}}$ and $R_{10^{-20}}'$ (in terms of, say, history-descriptions suggesting the existence of an `extended microstate' or a `microstate of the universe'  \citep[cf.][]{Hemmo-Shenker2022}). {\em If} we insist on the metaphysical project of saying what kind of computers exist, and {\em if} we rely on our best current theories of physics to tell us what kind of physical things exist, then the general relativistic computers are currently different kinds of physical things than the quantum-sized computers. In this exotic sense empirical structure physicalism is compatible with multiple realization (item 18 of Sect. \ref{section_comparison}; note that analogous comments would apply to a disambiguation of consciousness or of pain as a purely functional concept vs. a concept that can describe physically embodied agents).

\section{Closing remarks}\label{section_conclusion}

Empirical structure physicalism only claims supervenience of {\em current} special sciences on current physics. It is conceivable that in the future we may find new ways to describe and tell apart space-time regions or new empirical methods which are not available to current physics, and that such new descriptions and methods also become available for a future special science. For instance, in the future we may find an empirically adequate theory of quantum gravity. Given such a theory, it may also become possible to empirically adequately identify a space-time region which contains, say, a quantum gravitational computer. That is, it is possible that we develop a {\em future} computer science, understood as a future empirical theory whose formal structure also carries the same formal computer concept as does current computer science, but which is also capable to empirically adequately distinguish a space-time region with a quantum gravitational computer from a space-time region without one.

Although such a special science at a future time $t$ would not supervene on current physics, it would plausibly still supervene on future physics at time $t$. The reason is the same as explained in Sect. \ref{subsection_dualisms}: no measurement method will exist for a future special science at time $t$ whose different measurement outcomes would not manifest as differences in the empirical content of future physics at time $t$; in our example future physics will also include the future theory of quantum gravity. Hence we also have a good reason to believe that special sciences at a future time will supervene on physics at that future time.

Similar remarks apply to my reliance on the term `space-time region' to designate the locus of happenings of our world. For the purposes of this paper the term could be understood as a simple shorthand for `at a certain place for a certain amount of time', but if I were pressed to give a formal account of the latter idea, I would currently resort to submanifolds of the General Relativistic spacetime representing our actual world. A future theory of physics could, of course, understand a submanifold as an approximate concept that can be derived, under certain assumptions, from a future locus concept, and such a future locus concept may enable a more apt future articulation of empirical structure physicalism. Although such a future articulation may have relevance regarding the supervenience of some {\em future} special sciences, it does not invalidate the supervenience of {\em current} special sciences on current and future physics. This is because we can expect the notion of submanifold to supervene on the locus concept of a future theory of physics, analogously to how current physics supervenes on future physics (indeed, the ability to recover the spacetime of General Relativity in a limit is understood as an explicit constraint for any future theory of quantum gravity, see \citealt[298, 321]{Wutrich2017}). Thus, we have every reason to expect that even radical changes in the kinds of things future physics may suggest to exist will not radically alter the meaning of empirical structure physicalism in the future.

In this article I defined theory supervenience in an actualist context: by space-time region I meant a space-time region of our actual world. To relax this assumption, one may adopt a notion of possibility that coheres with a non-trivial, sensible concept of empirical adequacy. However, to do so by invoking the received view of nomic possibility (which is a notion of logical compatibility with the laws of a theory, see \citealt{Gyenis2025}) risks circularity. An alternative approach is to adopt the notion of empirical possibility of \cite{Norton2022}; details of such generalization shall be pursued elsewhere. 

What is, then, according to my proposal, `physical'? I have made no attempt to define this metaphysical concept. In my view, one thing we learned from the past decades of philosophy of physics is that it is extremely difficult to `read off' `the' ontology of a theory of physics. My justification for this view is complex, but independent of the pessimistic meta-induction or of Hempel's dilemma: the well-known cornucopia of interpretations of quantum mechanics and the problem of a metaphysical reading of gauge theories are examples of the many difficulties plaguing the metaphysical project. Thus, I believe that the picture of a peaceful division of labor, in which a theory of physics tells us what kinds of things exist and philosophers can simply accept these as physical things and run their metaphysical business with them is at best misleading, and will likely remain so until we find a fundamental theory of physics. I imagine philosophers of mind puzzlingly nodding along and saying: if this is right, then so much worse for {\em theory-oriented} metaphysical physicalism. However, since I do not believe in non-theory-oriented metaphysics, I draw the opposite conclusion: so much worse for the {\em metaphysical} project, at least currently. Suum cuique.

\section*{Acknowledgements}

The author would like to thank the audiences of the 2021 Physicalism mini-symposium and Workshop at the Research Centre for Humanities, the Theoretical Philosophy Forum at E\"otv\"os Univeristy, the MCMP Workshop: Physicalism, the LMU Research Seminar in Decision and Action Theory, and the Reduction and Inter-theoretical Relationships in the Sciences workshop for useful questions, and M\'arton G\"om\"ori, Gregor H\"orzer, John Norton, Elay Shech, and G\'abor Szab\'o for useful comments and suggestions on earlier versions of the manuscript. An earlier version of this article was published in Hungarian as \citep{Gyenis2022}. This research has been supported by the OTKA K-134275 grant.

%\bibliography{gyb_bibliography}

\begin{thebibliography}{}

\bibitem[\protect\citeauthoryear{Buzaglo}{Buzaglo}{2024}]{Buzaglo2024}
Buzaglo, D. (2024).
\newblock Two construals of Hempel's dilemma: a challenge to physicalism, not
  dualism.
\newblock {\em European Journal for Philosophy of Science\/}~{\em 14\/}(26),
  1--17.
\newblock \url{https://doi.org/10.1007/s13194-024-00590-9}.

\bibitem[\protect\citeauthoryear{Chalmers and McQueen}{Chalmers and
  McQueen}{2022}]{Chalmers-McQueen2022}
Chalmers, D.~J. and K.~J. McQueen (2022).
\newblock Consciousness and the collapse of the wave function.
\newblock In S.~Gao (Ed.), {\em Consciousness and Quantum Mechanics}, pp.\
  11--63. New York: Oxford Academic Press.
\newblock Quote based on their paper deposited as arXiv:2105.02314:
  \url{https://doi.org/10.48550/arXiv.2105.02314}.

\bibitem[\protect\citeauthoryear{Crane and Mellor}{Crane and
  Mellor}{1990}]{Crane-Mellor1990}
Crane, T. and D.~H. Mellor (1990).
\newblock There is no question of physicalism.
\newblock {\em Mind\/}~{\em 99}, 185--206.
\newblock \url{https://doi.org/10.1093/mind/XCIX.394.185}.

\bibitem[\protect\citeauthoryear{Earman}{Earman}{1989}]{Earman1989}
Earman, J. (1989).
\newblock {\em World enough and space-time}.
\newblock Cambridge: The MIT Press.

\bibitem[\protect\citeauthoryear{Fazekas, Gyenis, Hofer-Szab\'o, and
  Kert\'esz}{Fazekas et~al.}{2021}]{Fazekas-Gyenis-Szabo-Kertesz2021}
Fazekas, P., B.~Gyenis, G.~Hofer-Szab\'o, and G.~Kert\'esz (2021).
\newblock A dynamical systems approach to causality.
\newblock {\em Synthese\/}~{\em 198\/}(7), 6065--6087.
\newblock \url{https://doi.org/10.1007/s11229-019-02451-y}.

\bibitem[\protect\citeauthoryear{Firt, Hemmo, and Shenker}{Firt
  et~al.}{2022}]{Firt-Hemmo-Shenker2022}
Firt, E., M.~Hemmo, and O.~Shenker (2022).
\newblock Hempel's dilemma: not only for physicalism.
\newblock {\em International Studies in the Philosophy of Science\/}, 1--29.
\newblock \url{https://doi.org/10.1080/02698595.2022.2041969}.

\bibitem[\protect\citeauthoryear{Forrest}{Forrest}{2010}]{Forrest2010}
Forrest, P. (2010).
\newblock The identity of indiscernibles.
\newblock In E.~N. Zalta (Ed.), {\em The Stanford Encyclopedia of Philosophy}.
\newblock \url{https://plato.stanford.edu/entries/identity-indiscernible}
  (2022.06.26.).

\bibitem[\protect\citeauthoryear{Frigg}{Frigg}{2022}]{Frigg2022}
Frigg, R. (2022).
\newblock {\em Models and theories}.
\newblock London: Routledge.

\bibitem[\protect\citeauthoryear{G\"om\"ori}{G\"om\"ori}{2020}]{Gomori2020}
G\"om\"ori, M. (2020).
\newblock On the very idea of distant correlations.
\newblock {\em Foundations of Physics\/}~{\em 50}, 530--554.
\newblock \url{https://doi.org/10.1007/s10701-020-00332-w}.

\bibitem[\protect\citeauthoryear{Gyenis}{Gyenis}{2022}]{Gyenis2022}
Gyenis, B. (2022).
\newblock Elm\'elet-szuperveniencia fizikalizmus \'es a fizika j\"ov\H{o}beli
  v\'altoz\'asai.
\newblock {\em Magyar Filoz\'ofiai Szemle\/}~{\em 66\/}(2022/3), 41--66.
\newblock (in Hungarian).

\bibitem[\protect\citeauthoryear{Gyenis}{Gyenis}{2023}]{Gyenis2023}
Gyenis, B. (2023).
\newblock Az igazs\'ag harmadik pillanata.
\newblock {\em K\"ul\"onbs\'eg\/}~{\em 23\/}(1), 165--183.
\newblock \url{https://doi.org/10.14232/kulonbseg.2023.23.1.313}.

\bibitem[\protect\citeauthoryear{Gyenis}{Gyenis}{2025}]{Gyenis2025}
Gyenis, B. (2025).
\newblock Physical, empirical, and conditional inductive possibility.
\newblock {\em Philosophy of Physics\/}~{\em 3\/}(1), 4.
\newblock \url{https://doi.org/10.31389/pop.148}.

\bibitem[\protect\citeauthoryear{Halvorson}{Halvorson}{2019}]{Halvorson2019}
Halvorson, H. (2019).
\newblock {\em The logic in philosophy of science}.
\newblock Cambridge: Cambridge University Press.

\bibitem[\protect\citeauthoryear{Hemmo and Shenker}{Hemmo and
  Shenker}{2022}]{Hemmo-Shenker2022}
Hemmo, M. and O.~R. Shenker (2022).
\newblock Flat physicalism.
\newblock {\em Theoria\/}~{\em 88}, 743--764.

\bibitem[\protect\citeauthoryear{Hempel}{Hempel}{1969}]{Hempel1969}
Hempel, C.~G. (1969).
\newblock Reduction: ontological and linguistic facets.
\newblock In H.~Feigl and W.~Sellars (Eds.), {\em Readings in Philosophical
  Analysis}, pp.\  373--384. New York: Appleton Century Crofts.

\bibitem[\protect\citeauthoryear{Hempel}{Hempel}{1980}]{Hempel1980}
Hempel, C.~G. (1980).
\newblock Comments on Goodman's ways of worldmaking.
\newblock {\em Synthese\/}~{\em 45\/}(2), 193--199.
\newblock \url{https://doi.org/10.1007/BF00413558}.

\bibitem[\protect\citeauthoryear{H\"orzer}{H\"orzer}{2020}]{Horzer2020}
H\"orzer, G.~M. (2020).
\newblock {\em Understanding physicalism}.
\newblock Berlin/Boston: De Gruyter.

\bibitem[\protect\citeauthoryear{Ladyman}{Ladyman}{2023}]{Ladyman2023}
Ladyman, J. (2023).
\newblock Structural realism.
\newblock In E.~N. Zalta (Ed.), {\em The Stanford Encyclopedia of Philosophy\/}.
\newblock \url{https://plato.stanford.edu/entries/structural-realism/}
  (Summer 2023 ed.).

\bibitem[\protect\citeauthoryear{Lycan}{Lycan}{2003}]{Lycan2003}
Lycan, W. (2003).
\newblock Chomsky on the mind-body problem.
\newblock In L.~M. Anthony and N.~Horstein (Eds.), {\em Chomsky and his
  Critics}, pp.\  11--28. Oxford: Blackwell.

\bibitem[\protect\citeauthoryear{M\'arton}{M\'arton}{2019}]{Marton2019}
M\'arton, M. (2019).
\newblock Mi a ``fizikai''? k\'is\'erletek a ``fizikai'' fogalm\'anak
  meghat\'aroz\'as\'ara a kort\'ars fizikalizmusvit\'aban.
\newblock {\em Elpis\/}~{\em 2019\/}(1), 79--106.

\bibitem[\protect\citeauthoryear{Melnyk}{Melnyk}{1997}]{Melnyk1997}
Melnyk, A. (1997).
\newblock How to keep the `physical' in physicalism.
\newblock {\em Journal of Philosophy\/}~{\em 94\/}(12), 622--637.
\newblock \url{https://doi.org/10.2307/2564597}.

\bibitem[\protect\citeauthoryear{Nida-R\"umelin and Conaill}{Nida-R\"umelin and
  Conaill}{2019}]{NidaRumelin-Conaill2019}
Nida-R\"umelin, M. and D.~O. Conaill (2019).
\newblock Qualia: The knowledge argument.
\newblock In E.~N. Zalta (Ed.), {\em The Stanford Encyclopedia of Philosophy}.
\newblock \url{https://plato.stanford.edu/entries/qualia-knowledge}
  (2022.06.26.).

\bibitem[\protect\citeauthoryear{Norton}{Norton}{2003}]{Norton2003b}
Norton, J. (2003).
\newblock A material theory of induction.
\newblock {\em Philosophy of Science\/}~{\em 70\/}(4), 647--670.
\newblock \url{https://doi.org/10.1086/378858}.

\bibitem[\protect\citeauthoryear{Norton}{Norton}{2019}]{Norton2019}
Norton, J. (2019).
\newblock The hole argument.
\newblock In E.~N. Zalta (Ed.), {\em The Stanford Encyclopedia of Philosophy}.
\newblock \url{https://plato.stanford.edu/entries/spacetime-holearg}
  (2022.06.26.).

\bibitem[\protect\citeauthoryear{Norton}{Norton}{2022}]{Norton2022}
Norton, J.~D. (2022).
\newblock How to make possibility safe for empiricists.
\newblock In Y.~Ben-Menahem (Ed.), {\em Rethinking the Concept of Law of
  Nature}, Jerusalem Studies in Philosophy and History of Science. Springer,
  Cham.
\newblock \url{https://doi.org/10.1007/978-3-030-96775-8_5}.

\bibitem[\protect\citeauthoryear{Nye}{Nye}{1980}]{Nye1980}
Nye, M.~J. (1980).
\newblock N-rays: An episode in the history and psychology of science.
\newblock {\em Historical Studies in the Physical Sciences\/}~{\em 11\/}(1),
  125--156.
\newblock \url{https://doi.org/10.2307/27757473}.

\bibitem[\protect\citeauthoryear{Rabin}{Rabin}{2019}]{Rabin2019}
Rabin, S. (2019).
\newblock Nicolaus Copernicus.
\newblock In E.~N. Zalta (Ed.), {\em The Stanford Encyclopedia of Philosophy}.
\newblock \url{https://plato.stanford.edu/entries/copernicus} (2022.06.26.).

\bibitem[\protect\citeauthoryear{Shech}{Shech}{2019}]{Shech2019}
Shech, E. (2019).
\newblock Historical inductions meet the material theory.
\newblock {\em Philosophy of Science\/}~{\em 86\/}(5), 918--929.
\newblock \url{https://doi.org/10.1086/705524}.

\bibitem[\protect\citeauthoryear{Siegel}{Siegel}{2024}]{Siegel2024}
Siegel, S. (2024).
\newblock The contents of perception.
\newblock In E.~N. Zalta (Ed.), {\em The Stanford Encyclopedia of Philosophy\/}.
\newblock \url{https://plato.stanford.edu/archives/fall2024/entries/perception-contents/}
  (Fall 2024 ed.).

\bibitem[\protect\citeauthoryear{Smart}{Smart}{1978}]{Smart1978}
Smart, J.~J. (1978).
\newblock The content of physicalism.
\newblock {\em Philosophical Quarterly\/}~{\em 28\/}(113), 339--341.
\newblock \url{https://doi.org/10.2307/2219085}.

\bibitem[\protect\citeauthoryear{Stoljar}{Stoljar}{2021}]{Stoljar2021}
Stoljar, D. (2021).
\newblock Physicalism.
\newblock In E.~N. Zalta (Ed.), {\em The Stanford Encyclopedia of Philosophy}.
\newblock \url{https://plato.stanford.edu/entries/physicalism} (2022.06.26.).

\bibitem[\protect\citeauthoryear{Szab\'o}{Szab\'o}{2017}]{Szabo2017}
Szab\'o, L. (2017).
\newblock Meaning, truth, and physics.
\newblock In G.~Hofer-Szab\'o and L.~Wro\'nski (Eds.), {\em Making it Formally
  Explicit}. Cham: Springer.
\newblock \url{https://doi.org/10.1007/978-3-319-55486-0_9}.

\bibitem[\protect\citeauthoryear{Szab\'o}{Szab\'o}{2021}]{Szabo2021}
Szab\'o, L. (2021).
\newblock Physicalism without the idols of mathematics.
\newblock PhilSci Archive preprint,
  \url{http://philsci-archive.pitt.edu/18901}.

\bibitem[\protect\citeauthoryear{T\H{o}zs\'er}{T\H{o}zs\'er}{2023}]{Tozser2023}
T\H{o}zs\'er, J. (2023).
\newblock {\em The failure of philosophical knowledge}.
\newblock Bloomsbury: Bloomsbury Publishing.

\bibitem[\protect\citeauthoryear{van Fraassen}{van
  Fraassen}{1980}]{vanFraassen1980b}
van Fraassen, B.~C. (1980).
\newblock {\em The scientific image}.
\newblock Oxford: Oxford University Press.

\bibitem[\protect\citeauthoryear{Veronen}{Veronen}{1992}]{Veronen1992}
Veronen, V. (1992).
\newblock A weakness in Kuhn's regal argument.
\newblock {\em Science \& Technology Studies\/}~{\em 5\/}(1), 47--51.
\newblock \url{https://doi.org/10.23987/sts.55049}.

\bibitem[\protect\citeauthoryear{von Neumann}{von
  Neumann}{1932}]{vonNeumann1932}
von Neumann, J. (1932).
\newblock {\em Mathematical foundations of quantum mechanics}.
\newblock Princeton: Princeton University Press.
\newblock Translated by R. T. Beyer, 1955.

\bibitem[\protect\citeauthoryear{Votsis}{Votsis}{2011}]{Votsis2011}
Votsis, I. (2011).
\newblock Structural realism: continuity and its limits.
\newblock In A.~Bokulich and P.~Bokulich (Eds.), {\em Scientific
  Structuralism}. Dordrecht: Springer.

\bibitem[\protect\citeauthoryear{Wigner}{Wigner}{1962}]{Wigner1962}
Wigner, E. (1962).
\newblock Remarks on the mind-body question.
\newblock In I.~J. Good (Ed.), {\em The Scientist Speculates: An Anthology of
  Partly-baked Ideas}. London: William Heinemann Ltd.

\bibitem[\protect\citeauthoryear{Winther}{Winther}{2020}]{Winther2020}
Winther, R.~G. (2020).
\newblock The structure of scientific theories.
\newblock In E.~N. Zalta (Ed.), {\em The Stanford Encyclopedia of Philosophy\/}
\newblock
  \url{https://plato.stanford.edu/entries/structure-scientific-theories/}
  (Spring, 2021 ed.).

\bibitem[\protect\citeauthoryear{W\"utrich}{W\"utrich}{2017}]{Wutrich2017}
W\"utrich, C. (2017).
\newblock Raiders of the lost spacetime.
\newblock In D.~Lehmkuhl, G.~Schiemann, and E.~Scholz (Eds.), {\em Towards a
  Theory of Spacetime Theories}, Volume~13 of {\em Einstein Studies}. New York:
  Birkh\"auser.
\newblock \url{https://doi.org/10.1007/978-1-4939-3210-8_11}.

\end{thebibliography}

\end{document}